\documentclass{aa}

\usepackage{amsfonts}       
\usepackage{graphicx}
\usepackage{natbib}
\usepackage{textcomp}
\usepackage{physics}
\usepackage{xcolor}
\usepackage{soul}
\newcommand{\RomanNumeralCaps}[1]
    {\MakeUppercase{\romannumeral #1}}

\title{A solar coronal loop in a box: Energy generation and heating}
\author{C.~Breu\inst{\ref{inst1}}\and H.~Peter\inst{\ref{inst1}}\and R.~Cameron\inst{\ref{inst1}}\and S.K.~Solanki\inst{\ref{inst1} \and \ref{inst2}}\and D.~Przybylski\inst{\ref{inst1}}\and M.~Rempel\inst{\ref{inst3}}\and L.P.~Chitta\inst{\ref{inst1}}}
\authorrunning{Breu et al.}
\institute{Max Planck Institute for Solar System Research, Justus-von-Liebig-Weg 3, 37077 Göttingen, Germany\label{inst1}
\and School of Space Research, Kyung Hee University, Yongin, Gyeonggi 446-701, Republic of Korea\label{inst2}
\and
High Altitude Observatory, NCAR, P.O. Box 3000, Boulder, Colorado 80307, USA\label{inst3}}
\begin{document}

\abstract%
{Coronal loops are the basic building block of the upper solar atmosphere as seen in the extreme UV and X-rays. Comprehending how these are energized, structured, and evolve is key to understanding stellar coronae.}
{Here we investigate how the energy to heat the loop is generated by photospheric magneto-convection, transported into the upper atmosphere, and how the internal structure of a coronal magnetic loop forms.}
{In a 3D magnetohydrodynamics (MHD) model, we study an isolated coronal loop  rooted with both footpoints in a shallow layer within the convection zone  using the MURaM code. To resolve its internal structure, we limited the computational domain to a rectangular box containing a single coronal loop as a straightened magnetic flux tube.  Field-aligned heat conduction, gray radiative transfer in the photosphere and chromosphere, and optically thin radiative losses in the corona were taken into account. The footpoints were allowed to interact self-consistently with the granulation surrounding them.}
{The loop is heated by a Poynting flux that is self-consistently generated through small-scale motions within individual magnetic concentrations in the photosphere. Turbulence develops in the upper layers of the atmosphere as a response to the footpoint motions. We see little sign of heating by large-scale braiding of magnetic flux tubes from different photospheric concentrations at a given footpoint. The synthesized emission, as it would be observed by the Atmospheric Imaging Assembly (AIA) or the X-ray Telescope (XRT), reveals transient bright strands that form in response to the heating events. Overall, our model roughly reproduces the properties and evolution of the plasma as observed within (the substructures of) coronal loops.}
{With this model we can build a coherent picture of how the energy flux to heat the upper atmosphere is  generated near the solar surface and how this process drives and governs the heating and dynamics of a coronal loop.}

\keywords{Sun:corona, Sun:magnetic fields, Magnetohydrodynamics (MHD)}
\maketitle

\section{Introduction}
Solar coronal loops are bright structures of hot plasma confined by the magnetic field, observable in X-ray and extreme ultraviolet (EUV) light \citep{2014LRSP...11....4R}. Coronal loops can be found in the quiet sun as well as in active regions. The plasma temperatures in the coronal loops range from 1 to about 10 MK. The heating mechanism that sustains coronal loops is subject to active discussions.
Proposed models include wave heating \citep{1947MNRAS.107..211A, van_Ballegooijen_2011, 2012RSPTA.370.3217P, 2014ApJ...787...87V,2021ApJ...908..233S} and the braiding of the magnetic field lines by photospheric motions \citep{1972ApJ...174..499P,1983ApJ...264..642P,1988ApJ...330..474P,2002ApJ...576..533P,Rappazzo_2008}.

In the direct current (DC) model, small-scale horizontal photospheric motions at the loop footpoints lead to tangling of the magnetic field lines. The reconnection of braided field lines is thought to lead to small, impulsive heating events with energies of $\mathrm{10^{24}\; erg}$, the so-called nanoflares \citep{1972ApJ...174..499P,1983ApJ...264..642P,1988ApJ...330..474P}. Many such heating events could be capable of heating the corona to the observed temperatures.
The necessary energy inflow into the corona estimated from observations is $F=10^{7}\;\mathrm{erg\,cm}^{-2}\,\mathrm{s^{-1}}$ above an active region \citep{1977ARA&A..15..363W}, which leads to an estimate for the average heating rate of $Q=4\times 10^{-3}\; \mathrm{erg\,cm}^{-3}\,\mathrm{s}^{-1}$ \citep{van_Ballegooijen_2011}.

How energy injection and dissipation determines the internal structure of a loop is the subject of active research. Energy and mass transport predominantly occurs along the magnetic field. Instead of being a monolithic isolated structure, a loop potentially consists of a bundle of multithermal thinner flux tubes \citep{1994ApJ...422..381C}.
Imaging instruments such as the High-resolution Coronal Imager \citep[Hi-C;][]{2013Natur.493..501C,2019SoPh..294..174R} have yielded observations of fine threads in the loops. It has been suggested that the interior of coronal loops is dynamic and has a substructure at scales below the instrumental resolution \citep{1993ApJ...405..767G}. Theoretical considerations suggest that the individual threads could have widths down to 10 to 100~km \citep{2003SoPh..216...27B,2004ApJ...605..911C,2009A&A...499L...5V}. So far, Hi-C has provided the highest resolution in coronal observations and provided evidence of strands being as thin as 200 km \citep{2013ApJ...772L..19B,2020ApJ...892..134W}. Thick monolithic loops, however, have also been found in Hi-C observations, which do not show evidence for any substructure below 1.5 Mm \citep{2013A&A...556A.104P}.

The spatial resolution that can be achieved with numerical simulations based on magnetohydrodynamics (MHD) models covering a full active region is insufficient to fully take the internal driving inside a flux tube due to computational limitations into account. With grid spacings on the order of hundreds of kilometers, coronal structures are resolved by only a few grid points across their diameter \citep{2011A&A...530A.112B,2017ApJ...834...10R,2019A&A...624L..12W}, leading to less structured loops and potentially less energy input.  High resolution simulations are required to resolve the internal loop structure and study the heating mechanism in more detail.

Therefore, we simplify the loop geometry and straighten the coronal loop to fit into a Cartesian box. The coronal loop is modeled as a magnetic flux tube between two photospheric layers evolving independently. This is possible if the loop diameter is small compared to the radius of curvature of the loop, which is comparable to the loop length. In straightened loop simulations, coronal loops are represented as straightened magnetic flux bundles anchored in a layer within the convection zone at each footpoint. In simulations of a curved loop with a single photosphere, the actual loop to be modeled will fill only a small fraction of the coronal volume. This makes simulations with high resolution and long time-duration very costly in terms of computational time compared to our approach of a straightened loop in a box. This general type of stretched-loop model goes back to the seminal work of \citet{1996JGR...10113445G}.

Models of magnetic braiding and MHD turbulence in a straightened-out coronal loop in a rectangular box have been employed in a number of works.
Many of these impose a photospheric driver and then investigate its influence on the coronal structure and heating.
This is usually done by prescribing a velocity field driving the magnetic field line braiding at the boundary of the simulation domain. 
\citet{Rappazzo_2008, Rappazzo_2010, Rappazzo_2013} study braiding models in reduced MHD  with a prescribed velocity driver to mimic photospheric flows.
MHD turbulence is found to be responsible for the transport of energy to small scales \citep{Rappazzo_2008,2007ApJ...662..701B} where it is dissipated. These studies propose a complex relation between driving motions and resulting turbulent flows.

The first study of a straightened loop in full MHD was conducted by \citet{1996JGR...10113445G} and investigated the response of an initially homogeneous magnetic field to shearing motions at two boundaries. This has been complemented by including stratification in a 2D simulation \citep{2002A&A...383..685G}.

Instead of the energy transport into the corona, several studies of straightened loops focus on the energy transfer from the magnetic field to the plasma and investigate the turbulent relaxation of a braided initial magnetic field \citep{2010A&A...516A...5W,2011A&A...536A..67W,2015ApJ...805...47P,2016PPCF...58e4008P,2017ApJ...837..108P}. Local instabilities in flux tubes can trigger energy release by magnetic avalanches if the unstable flux tube disrupts neighboring threads \citep{2016ApJ...817....5H,2018A&A...615A..84R, 2020A&A...633A.158R}.

The realism of the photospheric driving in previous stretched-loop studies ranges from a simple shearing motion as employed in \citet{1996JGR...10113445G}, to  superposition of pulses recreating the observed coronal power spectrum used in \citet{Pagano_2019}. Both shearing and twisting motions are studied.
In \citet{2016ApJ...830...21R} the magnetic field is braided by random rotational motions. In these simulations of straightened loops, photospheric footpoint motions are assumed to occur on the length scale of granulation. \citet{van_Ballegooijen_2011} instead consider transverse motions within a magnetic concentration on a length scale smaller than the flux element to be responsible for coronal heating by Alfv\'{e}n wave turbulence. 
\citet{2006A&A...451.1101D,2006A&A...459..627D} look at the interaction between two flux tubes subjected to rotating and spinning motions, whereas \citet{van_Ballegooijen_2011} consider only a single flux concentration. Thus it does not take the splitting up and merging of flux elements that might play an important role for the energy supply into account. 

The cancellation of small-scale magnetic flux elements with a dominant main polarity and reconnection at the loop footpoints could play an important role for the energization and mass supply of coronal loops \citep{2017ApJS..229....4C,2018ApJ...862L..24P}, but has not been considered in any of these models. In particular, none of these models consider a realistic boundary condition at the photosphere, where magnetoconvection self-consistently drives the energization of the loop above.
In previous straightened-loop studies, the boundary of the simulation domain is located in the lower corona or chromosphere, not including the physics of the photosphere or convection zone where the actual driving takes place. Additionally, mass transport between photosphere, chromosphere, and corona and the interaction of magnetic elements and flux emergence at the surface is not considered.

We aim to study the internal structure of a loop and its connection to the photosphere, while having a self-consistent energy input into the corona arising from granular motion.
Here we carry out simulations in full 3D MHD.
A large fraction of previous studies of straightened coronal loops employ reduced MHD, which is not valid if the perturbation of the magnetic field becomes as strong as the guide field, which is the case if turbulence develops in the loop. In the framework of reduced MHD, the magnetic field is assumed to be a superposition of a strong magnetic guide field and a small perturbation perpendicular to the guide field. In contrast to reduced MHD, we solve the complete nonlinear system of MHD equations unrestricted by the assumption that the perpendicular component of the magnetic field is small compared to the guide field \citep{2017ApJ...839....2O}.\\
Our model includes the photosphere and near-surface convection zone. Including the upper layers of the convection zone in the simulation domain self-consistently leads to heating of the chromosphere and corona due to changing magnetic structures at the loop foot-points. In contrast to more idealized experiments studying one process in isolation, with a realistic driver for the magnetic field multiple heating processes are instead excited simultaneously. Observations point to coronal loops having a substructure below the instrument resolution, which cannot be resolved in models of a full active region. Such models can resolve only a few 100 km \citep{2019A&A...624L..12W}.

The structure of the paper is as follows.
We describe the methods and our loop model in detail in Sect.~\ref{section:mod}, analyze heat generation, transport, dissipation, and loop structure in Sect.~\ref{section:res}, discuss our results in Sect.~\ref{section:disc}, and present conclusions  in Sect.~\ref{section:conclusion}.

\section{Coronal loop model}
\label{section:mod}

In this section we describe our loop model, the code used to conduct the numerical experiments, the simulation setup, and employed initial conditions. In addition, we discuss the driving at the photosphere and the synthesis of the EUV and X-ray emission expected from the model. The loop is modeled in a simplified geometry as a straightened flux tube in a Cartesian box spanning the space between two shallow convection zone layers at its footpoints.

\subsection{Equations and loop model}
\label{section:num}

Radiative 3D MHD simulations are performed with the MURaM code \citep{2005A&A...429..335V}, using the extension of the code for coronal simulations by \citet{2017ApJ...834...10R}.
We solve the system of radiative magneto-hydrodynamic equations on a Cartesian grid in the form \citep{2014ApJ...789..132R,2017ApJ...834...10R}:
\begin{align}
\frac{\partial\rho}{\partial t}&=-\nabla \cdot(\rho \Vec{v}), \\
\frac{\partial \rho \Vec{v}}{\partial t}&=-\nabla\cdot(\rho \Vec{vv})-\nabla P +\rho g_{\mathrm{s}}(z)\hat{\Vec{z}}
+ \Vec{F}_{\mathrm{L}} + \nabla \cdot \tau\nonumber\\
&+\Vec{F}_{\mathrm{SR}},  \\
\frac{\partial E_{\mathrm{HD}}}{\partial t}&=-\nabla\cdot[\Vec{v}(E_{\mathrm{HD}}+P+q\Vec{B}/\abs{\Vec{B}})]
+\rho \Vec{v}\cdot   (g_{\mathrm{s}}(z)\hat{\Vec{z}})\nonumber \\
&+\Vec{v}\cdot \Vec{F}_{\mathrm{L}}
+\Vec{v}\cdot\Vec{F}_{\mathrm{SR}}+Q_{\mathrm{rad}}
+Q_{\mathrm{thin}}\nonumber\\
&+Q_{\mathrm{num,res}} + \nabla\cdot (\tau\cdot \Vec{v}),\\
\frac{\partial \Vec{B}}{\partial t}&=\nabla\times(\Vec{v}\times \Vec{B})+D_{\mathrm{num,res}},\\
\frac{\partial q}{\partial t}&=\frac{1}{\tau_{\mathrm{cond}}}(-f_{\mathrm{Sat}}\sigma T^{5/2}\Vec{B}/\abs{\Vec{B}}\cdot\nabla T-q).
\end{align}
The system of equations that is being solved consists of the equations for the conservation of mass, momentum, and energy as well as the induction equation and the equation for heat conduction.
$\rho$, $\Vec{v}$, $P$ and $\Vec{B}$ are mass density, velocity, pressure, and magnetic field, respectively. See Sect. \ref{subsection:geom} for a definition of the gravitational acceleration $g_{\mathrm{s}}(z)$ and discussion of the loop model. $\Vec{F}_{\mathrm{SR}}$ is a semi-relativistic correction term limiting the Alfv\'{e}n velocity in order to alleviate constraints on the timestep by using an artificially reduced speed of light \citep[see][]{2017ApJ...834...10R}. The Lorentz force $\Vec{F}_{\mathrm{L}}$ is computed as  $\Vec{F}_{\mathrm{L}}=\frac{f_{\mathrm{A}}}{4\pi}\nabla\cdot\left(\Vec{BB}-\frac{1}{2}\tens{I} \Vec{B}^{2}\right)
+(1-f_{\mathrm{A}})\frac{1}{4\pi}(\nabla \times \Vec{B} \times \Vec{B})$. $\tens{I}$ is the identity matrix and the prefactor $f_{\mathrm{A}}=1/\sqrt{1+\left(\frac{v_{\mathrm{A}}}{c}\right)}$ determines the transition between different treatments of the Lorentz force for high- and low beta regions to avoid spurious field-aligned components of the Lorentz force.
Instead of the total energy that would include also the magnetic energy, the plasma energy $E_{\mathrm{HD}}$ is used, which is the sum of internal and kinetic energy: $E_{\mathrm{HD}}=E_{\mathrm{int}}+0.5\rho v^{2}$. In order to maintain the $\nabla\cdot B=0$ condition, the code uses hyperbolic divergence cleaning \citep{2002JCoPh.175..645D}.

The model includes gray local thermodynamic equilibrium (LTE) radiative transfer in the photospheric and chromospheric layers as well as  Spitzer heat conduction along the magnetic field and optically thin radiative losses in the corona.
$Q_{\mathrm{rad}}$ is the radiative heating or cooling computed from LTE radiative transfer, while $Q_{\mathrm{thin}}$ denotes the optically thin radiative losses of the form $Q_{\mathrm{thin}}=-n_{\rm{e}}n_{\rm{H}}\Lambda(T)$, where $n_{e}$ is the electron density, $n_{\rm{H}}$ the number density of hydrogen nuclei, and $\Lambda(T)$ is a tabulated loss function \citep{2017ApJ...834...10R}. $q\cdot \Vec{B}/\abs{\Vec{B}}$ is the field-aligned conductive heat flux. 
$\tens{\tau}$ is the strain-rate tensor. The viscous force $\nabla\cdot\tau$ arising from nonzero viscosity is added in the momentum equation. The work done by the viscous term also leads to a contribution in the energy equation of $\Vec{v}\cdot(\nabla\cdot\tau)=\nabla\cdot(\tau\cdot\Vec{v})-\tau :(\nabla \Vec{v})$, where the second term on the right-hand-side is the energy taken out of the kinetic energy reservoir and added to the internal energy by viscous heating. Since the numerical scheme is conservative for the sum of kinetic and internal energy, the viscous heating term is not explicitly included in the energy equation. The energy removed from the kinetic energy reservoir by viscous heating is added to the internal energy reservoir and thus does not lead to a net change in the sum $E_{\mathrm{int}}+E_{\mathrm{kin}}$. 
 
Instead of explicit viscosity and a constant magnetic resistivity, only numerical resistivity and viscosity are taken into account in the simulations.The resistive heating is computed from the numerical fluxes at the cell interfaces for the magnetic field components. Instead of evaluating the strain-rate tensor, the viscous terms are calculated in a similar fashion from the fluxes for the velocity components. The numerical fluxes are calculated from a slope-limited diffusion scheme using a piecewise-linear reconstruction of the variables at the cell interfaces  \citep[for details on the diffusion scheme see][]{2014ApJ...789..132R,2017ApJ...834...10R}. The behavior of the diffusion is controlled by the parameter $h$, which, set to zero, will lead to a second order Lax-Friedrichs scheme, higher values concentrate the diffusion around monotonicity changes. Thus, the diffusivity of the scheme decreases with increasing $h$. For the $h$ parameter a value of $h$=2.0 was used in the convection zone and photosphere. In the corona, a value of $h$=1.25 was employed for the diffusion of mass, energy and momentum. A value of $h$=5.0 was used for the magnetic field. For a value of $h$ \textgreater 1 the diffusivity is switched off in sufficiently smooth regions and only kicks in around discontinuities in the solution  \citep{2014ApJ...789..132R,2017ApJ...834...10R}. This leads to a nonuniform resistivity in the simulation box. The resistive  heating term $Q_{\mathrm{num,res}}$ is then added in the energy equation to account for conservation of energy. The viscous heating is computed only for diagnostic purposes. However, there is an energy flux  due to numerical effective viscous force, which is nonzero but very small compared to the other terms. This viscous energy flux is added in the energy equation for consistency. The values we chose for the free parameter $h$ correspond to the high Prandtl number setting in \citet{2017ApJ...834...10R}. As a consequence, the viscous heating dominates over the resistive heating, especially in the chromosphere and low corona. In the coronal part, the viscous heating rate is roughly a factor of three higher than the resistive heating rate. 
To close the system of equations, we use an equilibrium ionization equation of state \citep{2017ApJ...834...10R}.

 In order to avoid time step constraints from the numerical treatment of the heat conduction and to speed up the simulation, hyperbolic heat conduction as described in \citet{2017ApJ...834...10R} is used to limit the maximum signal propagation speed. $q$ is the field-aligned heat flux, the parameter $f_{\mathrm{Sat}}$ is a factor taking into account the saturation of the conductive heat flux
 and $\sigma$ is the constant of the Spitzer heat conductivity, for which we use a value of $\sigma=10^{-6}\; \rm{erg}\,\rm{cm}^{-1}\,{\rm{s}}^{-1}\,{\rm{K}}^{-1}$. $\tau_{\mathrm{cond}}$ is a parameter chosen to determine the maximum propagation speed of the heat front.
 
 Likewise, we use the Boris correction in order to limit the Alfv\'{e}n speed to 3000 km/s to avoid very small timesteps \citep{2017ApJ...834...10R}. The Boris correction makes use of a semi-relativistic treatment of the MHD equations with an artificially reduced speed of light, leading to an asymptotic limit of the Alfv\'{e}n velocity.

The simulation box spans the solar atmosphere from the upper convection zone through the photosphere to the hot corona. The simulation domain is periodic in the x- and y-direction and has a horizontal extent of $\mathrm{6\;Mm}\times \mathrm{6\;Mm}$. At both ends of the simulation box in the s-direction, there is a photosphere and shallow convection zone driving the plasma evolution in the loop. The depth of the convection zone layer at each footpoint is 3.5 Mm and the extent of the box along the loop axis is 57 Mm, leading to an effective loop length of 50 Mm. Due to its length, it can be assumed that footpoints of the loop at both ends are far apart (about 30 Mm in case of a semi-circular loop of the same length).
The simulation domain is covered by $100\times 100\times 950$ gridpoints, giving a spatial resolution of 60 km. The grid is equidistant. The current version of MURaM does not allow for mesh refinement to better resolve the transition region or photosphere. In later studies this resolution will be significantly increased.

\subsection{Loop geometry}
\label{subsection:geom}

\begin{figure} 
\resizebox{\hsize}{!}{\includegraphics{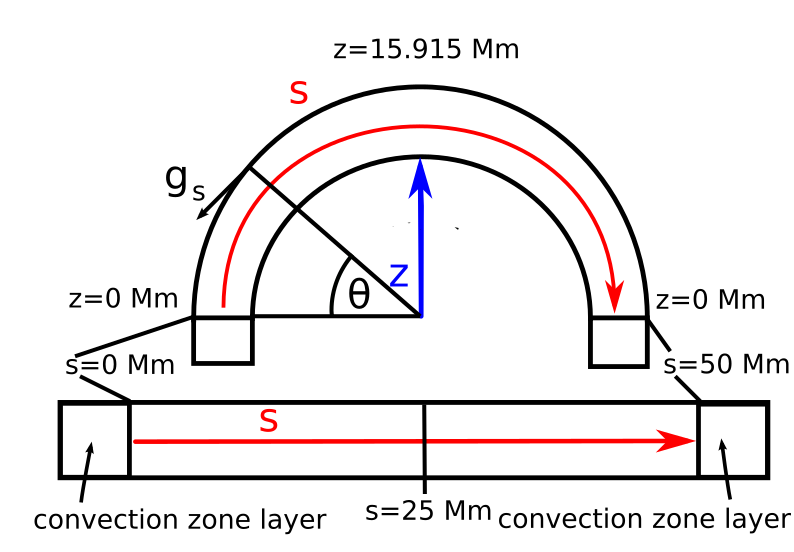}}
\caption{Sketch to illustrate the loop geometry.}
\label{fig:sketch}
\end{figure}

The MURaM code was modified to allow for a simplified loop geometry. Here we define $z$ as the geometrical height above the photosphere and $s$ as the coordinate along the loop axis. The gravitational acceleration was modified above the photospheric layer to  account for the curved geometry of the loop assuming a semicircular shape. Only the component along the loop axis, here in the s-direction was considered, 
\begin{equation}
g_{\mathrm{s}}(z) = g\cdot\cos\left(\pi\frac{s(z)-h_{\mathrm{photo}}}{s_{\mathrm{max}}-2\cdot h_{\mathrm{photo}}}\right).
\end{equation}
Here we assumed a semi-circular loop with straight vertical ends. The gravitational acceleration was modified above the photosphere, below the solar surface a constant value was assumed. Here, while $h_{\mathrm{photo}}$ is the height of the photosphere measured from the bottom boundary of the simulation box, which we set to 3.5 $\mathrm{Mm}$. The coordinate $s(z)=\theta\times(\pi/180\cdot R)$ is the arclength of the loop with $R$ being the loop radius. With a height of the convection zone layer of 3.5 Mm and a loop length of 50 Mm, we have a loop radius of ca. 15.915 Mm.\\
$s_{\mathrm{max}}$ is the total extent of the simulation box in the direction along the loop axis, which is 57 $\mathrm{Mm}$ in our case. The setup and the relation between the height z and the coordinate s along the loop axis are illustrated in Fig. \ref{fig:sketch}.
The component of the gravitational force perpendicular to the loop axis can be neglected if the gravitational force is negligible compared to the Lorentz force. This condition is fulfilled for regions with low plasma-beta, defined as the ratio of gas pressure to magnetic pressure $\beta = \frac{p_{\mathrm{gas}}}{\mathrm{p_{mag}}}$. The condition $\beta \ll 1$ is fulfilled for the coronal part of the simulation box.

Both boundaries at the loop footpoints are located in the near-surface convection zone about 3.5 Mm below optical depth unity. At these boundaries we allow for mass flux across the boundary layer so that the shallow convection zone can develop below the surface. The simulated layers of the bottom of these shallow convection zone layers at both ends of the loop are open for outflows (downward directed). The entropy and pressure of the inflows through the bottom boundaries is prescribed in order to fix the mass contained in the box and the brightness of the solar surface. 

Further adjustments were made to the radiative transfer computation. The optical depth was integrated from both sides of the loop from the boundary in the convection zone upward in the direction of the loop apex. The incoming radiation from each side was set to zero in the midplane at the loop apex to ensure that the radiation coming from the photosphere on one side does not heat the opposite photospheric layer. In the coronal part of the computational domain, only optically thin radiative losses play a role and the plasma is not affected by the incoming radiation from both footpoints.

\subsection{Initial conditions}
\label{sect:init}

As a first step toward an initial condition we used a stratified atmosphere in hydrostatic equilibrium. We started from a shallow convection simulation without a corona, which extends only $1\;\mathrm{Mm}$ above the surface. We define the location of the photosphere as the height at which the horizontally averaged optical depth is unity. 
The shallow simulation box was evolved without a magnetic field until convection reaches a statistically steady state. Subsequently, a uniform axial magnetic field of 30 G was added and the box was evolved for an hour. This corresponds to many convective turnover times and several Alfv\'{e}n crossing times. This should ensure that the solution becomes independent of the initial condition. During this time, the magnetic field is advected into the intergranular lanes due to convective motions. The magnetic field is concentrated in flux tubes of kilogauss strength. The average unsigned surface magnetic field is amplified by the convection to roughly 70 G, which corresponds to a weak plage region.

To get the initial condition for the coronal loop model, we took a shallow box of near-surface magnetoconvection, duplicated it, flipped the polarity, and stitched these two boxes together with a long coronal part.
The shallow rectangular box formed the loop footpoints at both boundaries and a hot corona was added between the two photospheric layers. Before that, a small random velocity was added to each gridpoint for one of the shallow boxes to ensure an independent evolution of the plasma in the box and therefore of the magnetic field at each footpoint.
As initial condition in the coronal part, we chose a plane parallel atmosphere.
We prescribed an initial temperature profile following a hyperbolic tangent in each loop leg with a maximum temperature of 1 MK at the loop apex.
The temperature profile is symmetric to the loop midplane. Density and pressure were then calculated from the temperature profile under the assumption of hydrostatic equilibrium.
The setup was evolved for another hour in order to become independent of the initial condition.\\

Some important effects have been neglected in this setup, such as the expansion of the loop with height above the chromosphere. A study of the effect of loop geometry on heating is conducted for 1D models in \citet{2013ApJ...773...94M}. For loop models with nonuniform cross-section, the authors find a higher density at the apex and a higher pressure at the coronal base, which leads to an increased emission in EUV compared to the model with a uniform cross-section. This is due to an increased maximum heat flux into the loop for the nonuniform cross-section models leading to larger radiative losses in the transition region. Since the temperature at the apex is roughly the same for both the uniform and the nonuniform area models, the density at the apex is enhanced for the expanding loop models. Therefore, we expect our model to underestimate the loop density and hence the coronal emission.

\subsection{Magnetoconvection in the bottom layers}

\begin{figure*} 
\resizebox{\hsize}{!}{\includegraphics{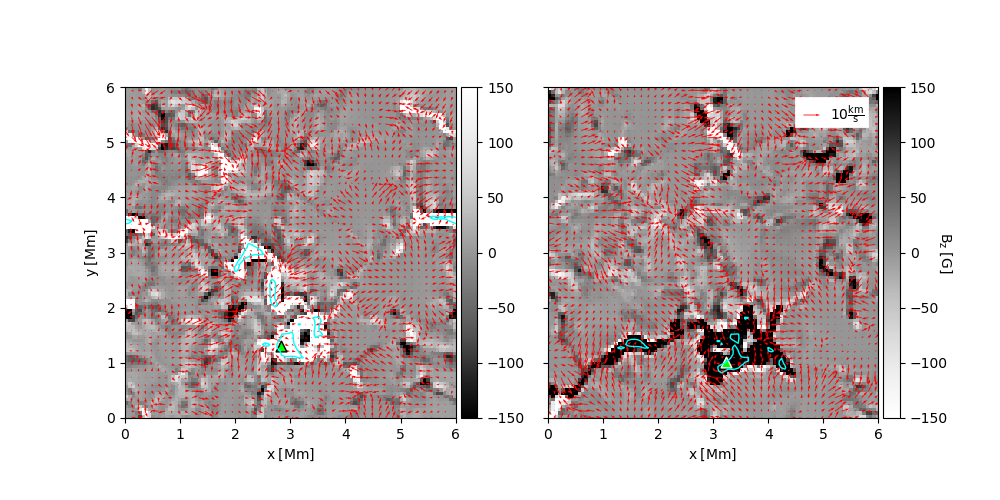}}
\caption{Vertical magnetic field at the photosphere at both loop footpoints at the depth where $\langle \tau \rangle = 1$. The snapshot was taken at $t=24.34\; \mathrm{min}$. The red arrows show the magnitude of the horizontal velocity at the photospheric level (see legend at top right). The blue contours indicate regions with $\lvert B_{\mathrm{z}}\rvert >1000\; G$. The green triangles mark the footpoints of the fieldline investigated in Sect. \ref{section:fline}. For a discussion of the initial condition for the magnetic field see Sect. \ref{sect:init}. A movie is available online.}
\label{fig:photo}
\end{figure*}

The simulation is driven by magnetoconvection.
Fig. \ref{fig:photo} shows the vertical magnetic field $B_{\mathrm{z}}$ at unity optical depth at 500 nm, $\langle \tau\rangle=1$, for both photospheric layers at each footpoint of the coronal loop. Magnetoconvection concentrates the magnetic field in flux tubes at the vortices of granular downflow lanes \citep{2014ApJ...789..132R}.
These magnetic elements with kilogauss field strengths have widths of several tens to hundreds of kilometers. Blue contours in Fig. \ref{fig:photo} highlight several flux tubes with $\lvert B_{\mathrm{z}} \rvert >1000\; G$. 

In addition to displacement of the flux concentrations by convective motions, transverse motions occur within flux elements on length scales smaller than the flux element \citep{van_Ballegooijen_2011,2012A&A...541A..68M,2020ApJ...894L..17Y,2021A&A...645A...3Y}. Observations do point to such internal motions \citep{2012ApJ...752...48C}. This is possible because the flux tubes are surrounded by turbulent convective downflows. Flux elements are not only displaced as in the flux tube tectonics model by \citet{2002ApJ...576..533P}, but also deformed and field lines rooted within them are intermixed. Fig. \ref{fig:photo} is supplemented with a movie showing the time evolution of the photospheric magnetic field.
Small-scale footpoint motions within the kG flux tubes have an rms velocity around 2 $\mathrm{km \, s^{-1}}$.
The rms velocity of photospheric motions away from flux concentrations is larger, about 3  $\mathrm{km \, s^{-1}}$. The internal motions within a flux tube are comparable to the setup by \citet{van_Ballegooijen_2011}.
Instead of being prescribed at the boundaries, the driving of each loop footpoint at the coronal base thus arises self-consistently from magnetoconvection.

\subsection{Synthetic emission}

\label{section:em}

In order to compare our loop model to observational data, we computed synthetic observations as would be seen in EUV and X-ray emission. To characterize emission from the corona around 1 to 2 MK we synthesized the EUV channels of the Atmospheric Imaging Assembly \citep[AIA;][]{2012SoPh..275...17L}. For the synthesis of X-ray emission we used the response for the Al-Poly filter of the X-Ray Telescope for the Hinode Mission \cite[XRT;][]{2007SoPh..243...63G}. 
The emission is dominated by collisionally excited lines. In case of XRT there is also a contribution from the X-ray continuum. Thus the emission is proportional to the product of electron and hydrogen number density and the energy loss per time and unit volume is given as
\begin{equation}
    \varepsilon=n_{e}n_{H}K(T),
\end{equation}
where $K(T)$ is a kernel given by the contribution function of the lines and continua in the bandpass accounting for the effective area of the instrument. These kernels for AIA are given in \citet{2012SoPh..275...41B}, and \citet{2012SoPh..275...17L}. For XRT, the kernels for the different filters are given in \citet{2007SoPh..243...63G}. Here we used these kernels as they are provided through the SolarSoft\footnote{https://www.lmsal.com/solarsoft/} package for AIA and XRT.
We computed the emission from the temperature and density in each gridpoint of the output data.
The synthetic observations were then obtained by integrating along the line of sight perpendicular to the loop axis:
\begin{equation}
 F=\int n_{\rm{e}}n_{\rm{H}}K(T)dx.
\end{equation}

\section{Results}
\label{section:res}

In the following sections, we discuss the general properties and time evolution of the loop, the injection of energy into the upper atmosphere, its dissipation, and the resulting overall structure and dynamics of the loop. We also synthesize observables from the simulation data.

\subsection{Overall behavior}
\label{subsection:overall}

\begin{figure*} 
\centering
\resizebox{.9\hsize}{!}{\includegraphics{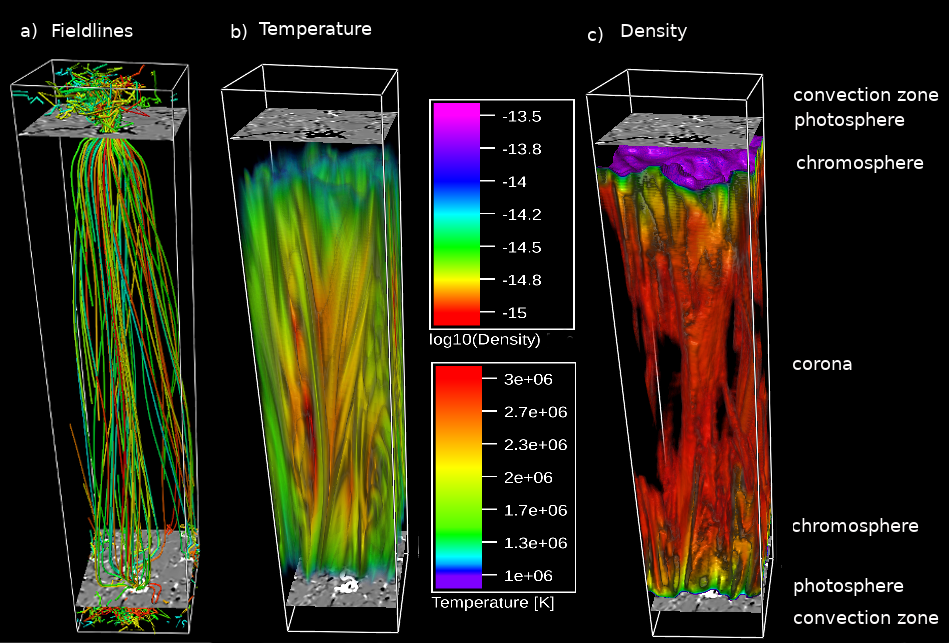}}
\caption{Simulation setup. (a) Example magnetic field lines in the simulation box traced from locations with strong magnetic field. The vertical magnetic field is plotted on a horizontal cut at the average photospheric height. (b) Volume rendering of the loop temperature in K. (c) Volume rendering of the loop density in $\mathrm{g cm^{-3}}$. The s-axis (along the loop) has been compressed by a factor of two for better visibility.}
\label{fig:twisting}
\end{figure*} 

\begin{figure*} 
\sidecaption
\includegraphics[width=.6\textwidth]{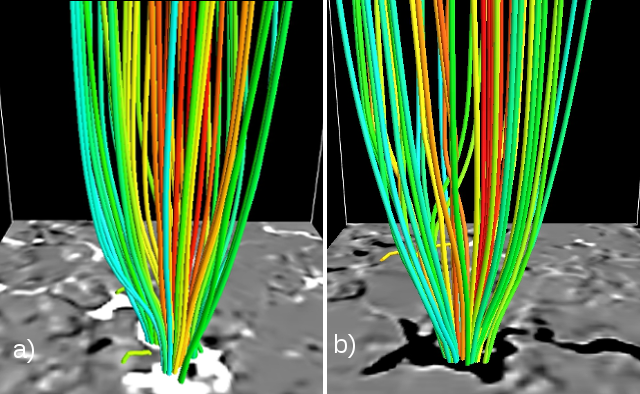}
\caption{Close-up of the footpoints of the loop shown in Fig. \ref{fig:twisting}. Magnetic field lines are traced from regions of enhanced magnetic field at a height of 2 Mm above the photosphere. The colors show the fieldline seed index.}
\label{fig:footpoints}
\end{figure*}
The simulation setup is illustrated in Fig. \ref{fig:twisting}. The snapshot displayed is the same as in Fig. \ref{fig:photo}. The vertical component of the magnetic field is plotted in two cuts at the heights of each photosphere. 
Magnetic field lines were traced from the location of strong photospheric magnetic field. The field line bundles reveal an internal twist of the loop caused by photospheric motions. Panels b) and c) show a volume rendering of density and temperature, respectively. 
A contiguous substructure with locally enhanced temperature and density develops in the coronal part of the simulation box and is aligned with the magnetic field.
A close-up of the loop footpoints is shown in Fig \ref{fig:footpoints}. The fieldlines were traced from locations with increased magnetic field in the lower corona to the photosphere. The magnetic field is mainly rooted in a few strong magnetic elements. At both footpoints, the magnetic field lines show signs of internal twisting and braiding within a single magnetic concentration.

\begin{figure} 
\resizebox{\hsize}{!}{\includegraphics{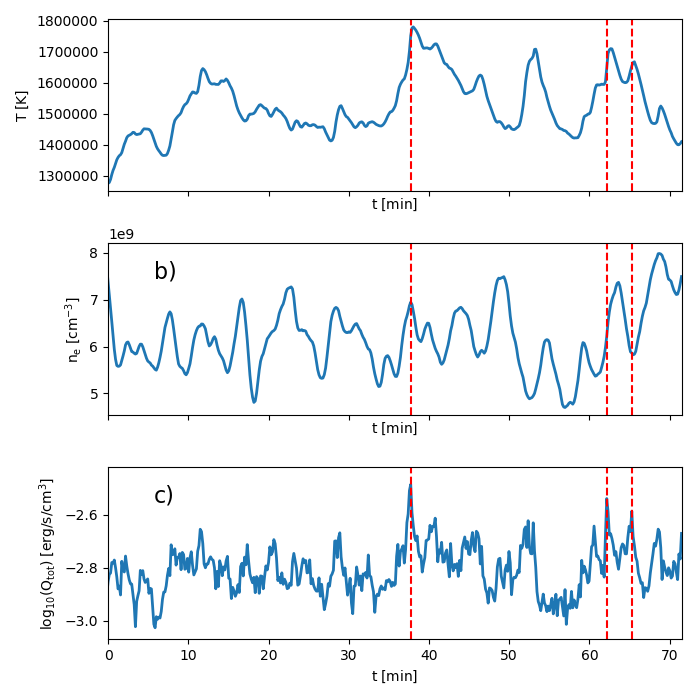}}
\caption{Averaged quantities in the coronal part of the simulation domain as a function of time. We define t=0 s as the time when we start collecting snapshots after the simulation extended to the coronal part has run for an hour of solar time. Panel (a) shows the temperature, panel (b) the electron density and panel (c) the sum of viscous and resistive heating. All quantities have been averaged over regions with densities below $\mathrm{10^{-12}\; g/cm^{3}}$. The vertical dashed lines highlight several strong heating events. For a discussion of the time evolution see Sect. \ref{subsection:overall} and \ref{subsection:poynt_heat}.}
\label{fig:avg}
\end{figure} 

\begin{figure} 
\resizebox{\hsize}{!}{\includegraphics{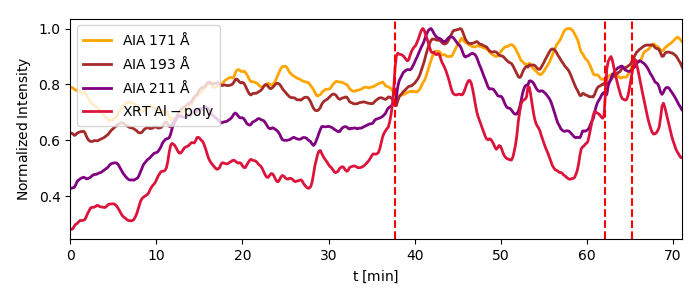}}
\caption{Normalized intensity integrated over the coronal part of the simulation domain as a function of time. The vertical dashed lines highlight the same strong heating events as in Fig. \ref{fig:avg}. For a discussion of the time evolution see Sect. \ref{subsection:overall} and \ref{subsection:poynt_heat}.}
\label{fig:aia_time_evo}
\end{figure} 

The time evolution of the averages of coronal temperature, electron density, and heating rate is displayed in Fig. \ref{fig:avg}. All quantities were averaged over the coronal part of the computational domain, which is defined here as the part of the simulation domain with densities below $\mathrm{10^{-12}\; g/cm^{3}}$ or particle densities below $\mathrm{5.98\times 10^{11}\; cm^{-3}}$. All quantities show oscillatory behavior with periods of several minutes.
The total heating rate, which is the sum of viscous and resistive heating, shows several strong heating events at $\mathrm{37.7\; min}$, $\mathrm{62.2\; min}$ and $\mathrm{65.3\; min}$. The heating rate is intermittent, energy is injected in pulses of varying duration. Longer heating events with lifetimes of five to ten minutes are superposed with short spikes of a duration on the order of seconds to minutes.

This is qualitatively consistent with the results of \citet{2013A&A...550A..30B} for a global active region simulation, where they find heating transient in time and space with pulses having a duration of a few minutes up to half an hour. Their study, however, is conducted with a lower resolution of 390  km in the x- and y-direction and 240 km in the z-direction. 

Bursty heating, which occurs on shorter timescales than the temperature evolution, is also found by \citet{2020A&A...633A.158R} for a simulation of an MHD avalanche in three twisted magnetic field threads  triggered by a local instability. Once the system reaches a continuously driven state, it continues to generate short bursts of heating superposed on a continuous background heating rate.

Over a simulation time of one hour, the mean temperature varies from 1.3 to 1.8 MK. 
We find mean electron densities on the order of $\mathrm{5\times10^{9}\; cm^{-3}}$ in the corona. After the heating events at $\mathrm{37.7\; min}$ and $\mathrm{62.2\; min}$, the coronal temperature shows a steep increase. Between $\mathrm{60\; min}$ and $\mathrm{70\; min}$ the electron density increases roughly by a factor of 1.5, reaching a peak with a delay of five minutes after the temperature peak at $\mathrm{62.2\; min}$. The mean temperature reaches its maximum value at $\mathrm{38\; min}$ and decreases to a local minimum value over a timescale of a quarter of an hour.\\

In response to an increase in temperature and density, the loop brightens in EUV and X-ray wavelengths. A time evolution of the normalized emission in the coronal part of the loop in the 171, 193, and 211 \AA\ bands as well as in the Al-poly filter of XRT is shown in Fig. \ref{fig:aia_time_evo}. After the heating event at $\mathrm{37.7\; min}$, the loop brightens up in the X-ray band first. As the loop cools, the increase in X-ray emission is followed by peaks in the 211, the 193, and the 171 \AA\; channel, which are sensitive to emission from plasma at subsequently lower temperatures. Further peaks in the emission appear after the heating events at $\mathrm{52\; min}$, $\mathrm{62.2\; min}$, and $\mathrm{65.3\; min}$.

\subsection{Poynting flux and heating}

In this section we discuss the energy input and heating. We start with briefly outlining the field line tracing method and follow this with a discussion of energy injection as well as dissipation along an individual field line in time. We then present a description of the spatial distribution of the injection of Poynting flux into the loop, the resulting loop structure and the development of current sheets.

\begin{figure*} 
\resizebox{\hsize}{!}{\includegraphics{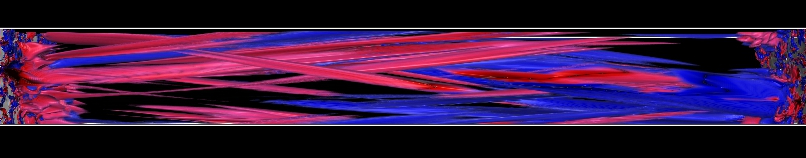}}
\caption{Volume rendering of the axial component of the Poynting flux in the Solar atmosphere in a given snapshot at time $24.34\; \mathrm{min}$. The Poynting flux directed in the positive $s$-direction (that is, from left to right in this figure) is shown in red and the Poynting flux directed in the negative $s$-direction is shown in blue. The $s$-axis is defined as the coordinate along the loop axis. The range of the color scale is $\mathrm{-1\times 10^{8}}$ to $+\mathrm{1\times10^{8}\; erg\;cm^{2}\;s^{-1}}$. The figure shows the volume encompassing chromosphere and corona between the two photospheres, one at each footpoint. See Sect. \ref{section:struct}.}
\label{fig:poynt}
\end{figure*}

\begin{figure*} 
\resizebox{\hsize}{!}{\includegraphics{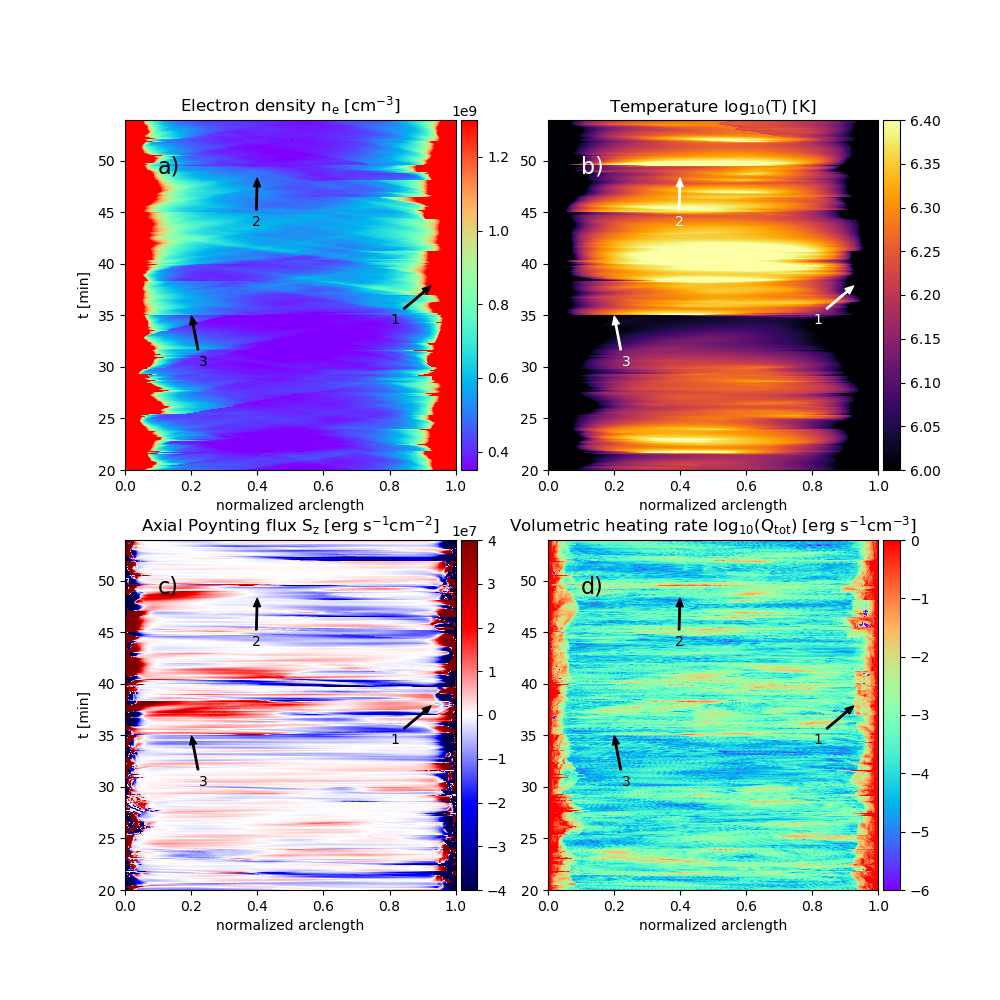}}
\caption{Spatio-temporal evolution along a field line with the arclength of the field line on the abscissa and time on the ordinate. The quantities shown are electron density (a), temperature (b), axial component of the Poynting flux (c), and total heating rate (d). Temperature and heating rate are scaled logarithmically. Positive Poynting flux (red) indicates energy propagating in the direction of increasing arc length (to the right), negative Poynting flux (blue) is in the opposite direction. The seed point from which the field line is traced is advected in time with the plasma velocity. The arrows mark locations of Poynting flux reaching the opposite chromosphere (1), dissipation (2), and reconnection (3), see Sect. \ref{section:fline}. For clarity, only the part of the time series containing the strongest heating event at 37.7 min depicted in Fig. \ref{fig:avg} is shown.}
\label{fig:spacetime}
\end{figure*}

\begin{figure*} 
\sidecaption
\resizebox{12cm}{!}{\includegraphics{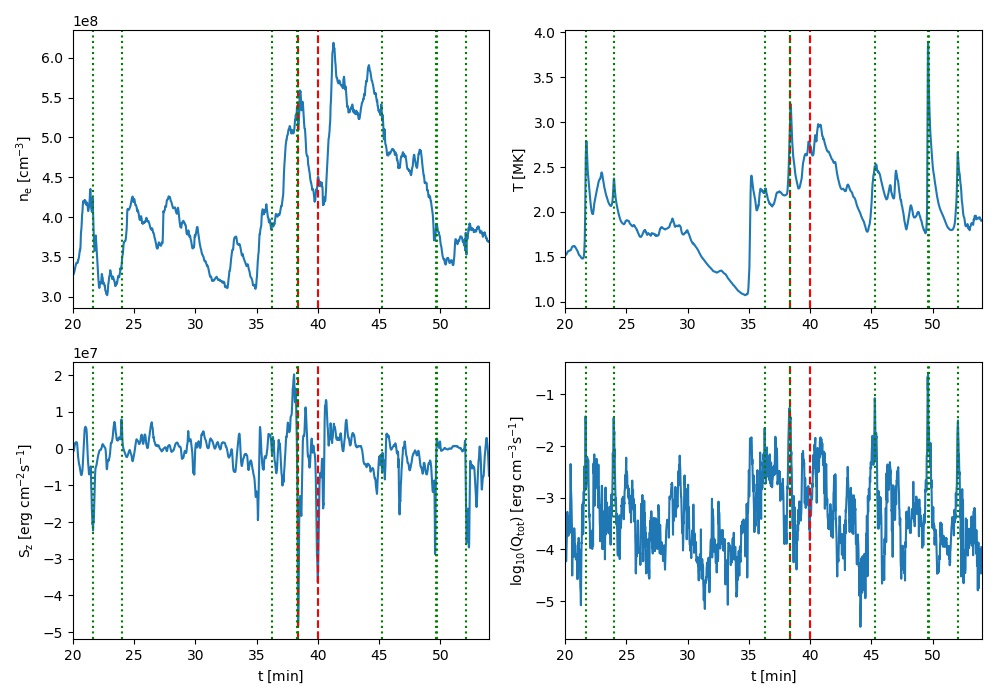}}
\caption{Time evolution of various quantities at the intersection of the field line shown in Fig. \ref{fig:spacetime} with the loop apex. The red dashed lines mark the two strongest Poynting flux events, while the green dotted lines mark the six strongest heating events, see Sect. \ref{section:fline}. For clarity, only part of the time series containing the strongest heating event at 37.7 min depicted in Fig. \ref{fig:avg} is shown.}
\label{fig:apex}
\end{figure*}

 \subsubsection{Field line tracing}
\label{section:ftrace}

In order to relate heating events to energy influx from the photosphere, we followed a set of magnetic field lines in time and space.
The field lines were integrated in space from randomly chosen seed points at the apex using a 4th order Runge-Kutta algorithm. We followed each field line in time by assuming that the magnetic field and the plasma are frozen together because of the high electric conductivity. The seed point chosen to lie on the midplane of the loop (that is, the loop apex) was advected with the velocity field of the plasma. The updated field line at the next time step was traced from the updated location of the seed point, then the intersection of the new field line with the midplane of the simulation box was used as the new seed point to be advected in the subsequent timestep. 
To show results for one sample fieldline, we selected an individual field line connected to plasma reaching a temperature of several million Kelvin at the loop apex. This field line was found to be anchored in the kilogauss magnetic concentrations at each loop footpoint.

\subsubsection{Behavior of an individual field line}
\label{section:fline}

We now investigate one selected field line to illustrate the changes along field lines in time.
The temperature at the apex of this field line reaches values of up to 4.0 MK, thus the selected field line is connected to one of the hottest strands within the loop.
The Poynting flux to heat this strand originates from both footpoints. The preferred direction of the Poynting flux at each footpoint is upward into the loop (red on the left side, blue on the right side in Fig. \ref{fig:poynt}).
The values of various coronal quantities interpolated along the selected field line in form of a spacetime diagram as a function of arclength along the field line and time over a time range of 35 minutes are displayed in Fig. \ref{fig:spacetime}.

The Poynting flux in the loop shows periodic behavior. While this period is difficult to see by eye in Fig 8c, or in Fig 9, a Fourier analysis of the (mean) axial Poynting flux shows a clear enhancement of power for periods between 30 s and 50 s. This timescale corresponds to the Alfv\'en crossing time through the coronal part of the loop.
The spacetime diagram shows bursty intermittent heating distributed along the field line as a response to the energy influx.
Between $t=35$ min and $t=55$ min a number of strong short-lived heating events occur, which cause the field line to heat up. A clear increase in temperature is seen in panel (b) after $t=35$ min.
Following the onset of the heating, the electron density shows filling of the loop with plasma.

The evolution of the same quantities at the apex of the selected field line is shown in Fig. \ref{fig:apex}. The Poynting flux in the selected time range stems predominantly from the left footpoint. While the temperature reaches its maximum between 17 and 20 minutes and subsequently decreases, the electron density continues to rise in response to the upflows from the footpoints. The two strongest Poynting flux events are associated with a dip in the electron density and subsequent increase. Both are closely followed by spikes in the total heating rate. While especially longer heating events lead to higher temperatures, some strong but short events are not associated with an increase in temperature, for example the first event marked by a vertical line in Fig. \ref{fig:apex}.

It is not straightforward to determine a preferred location of energy deposition in the corona. Regions of the highest level of volumetric heating are concentrated mainly near the footpoints in the chromospheric parts of the loop. The coronal part of the field line shows several strong isolated heating events, but moderate heating occurs along the whole length of the field line. Overall, there seems to be no preferred spatial location for heating in the corona apart from a slight increase in the number of strong localized heating peaks in the upper parts of the loop, which is consistent with \citet{2020A&A...633A.158R}.

We find instances at which the Poynting flux injected at one footpoint reaches the chromospheric layer at the other footpoint of the loop, for example at $t=38$ min (see arrow 1 in Fig. \ref{fig:spacetime}), but at the same time find evidence for the Poynting flux being dissipated in the atmosphere before reaching the opposite footpoint of the loop, thereby heating the plasma.  One example is shown at $t=48$ min where Poynting flux is injected into the loop from both sides and dissipated near the loop apex, leading to a strong heating event (see arrow 2). Reflection of Poynting flux at the opposite transition region due to the steep gradient of the Alfv\'{e}n velocity in the lower atmosphere cannot be excluded. 
We do not see possible reflection events every time the Poynting flux reaches the chromosphere at the opposite side of the loop, instead the Poynting flux seems to be absorbed by the denser material in the lower atmosphere in most cases.

Several reconnection events take place along the fieldline during the time covered in Fig. \ref{fig:spacetime}.
In the time range from $t=35$ min to 50 min, when the highest temperatures and the strongest Poynting flux input along the field lines are observed, the field line is connected to the interior or the edges of the kilogauss flux tubes where plasma is pushed into the downflow lanes at least at one footpoint. The largest Poynting flux is seen for a combination of the field line being connected to a region with high magnetic field and strong horizontal flows at the photospheric level.

It is not trivial to follow the field lines in time because they do not  retain their identity over a long time range. Similar to the simulation of \citet{2018A&A...615A..84R}, the field line undergoes frequent reconnection events with neighboring field lines. Furthermore, in the numerical model the frozen-in condition is not strictly fulfilled due to a finite resistivity, so that the field line can slip through the plasma. Due to numerical inaccuracies, the field line tracing algorithm can jump to a neighboring fieldline. The algorithm becomes structurally unstable near reconnection regions, where field lines cannot be clearly defined. 
The arrow (3) in Fig. \ref{fig:spacetime} at time t=35.1 min  indicates a possible reconnection event identified by discontinuities in density and a strong increase in heating rate along the field line.

\subsection{Loop structure}
\label{section:struct}

\begin{figure*} 
\resizebox{\hsize}{!}{\includegraphics{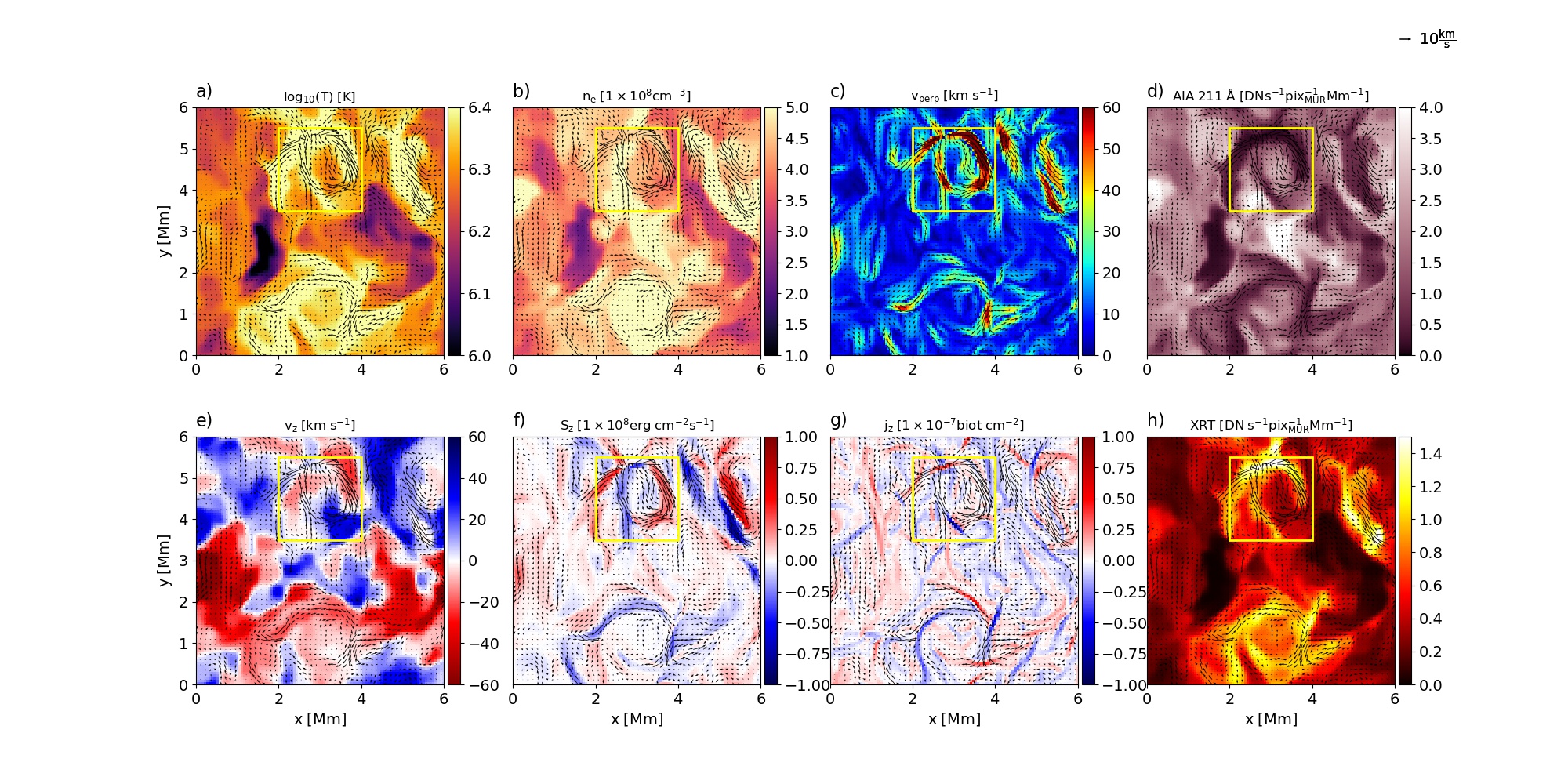}}
\caption{Cut through the loop at the apex (cross section perpendicular to the loop axis). (a) Temperature; (b) Electron density; (c) Magnitude of the velocity perpendicular to the loop axis; (d) emission in the 211 \AA\; channel of AIA, (e) Axial component of the velocity; (f) Axial component of the Poynting flux; (g) Axial component of the current density, (h) emission as seen with the Al-poly filter of XRT. The black arrows illustrate the horizontal velocity field. The yellow box highlights the location of a vortex with enhanced temperature and density. The snapshot was taken at $\mathrm{t=41.02\;min}$. See Sect. \ref{section:struct}. }
\label{fig:swirl_cuts}
\end{figure*}

\begin{figure*} 
\resizebox{\hsize}{!}{\includegraphics{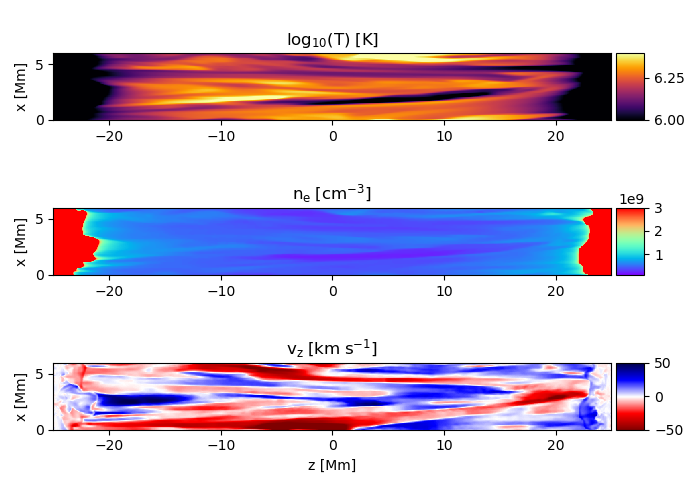}}
\caption{Axial cut through the simulation box. From top to bottom: Temperature, electron density, axial component of the velocity. For a discussion see \ref{section:struct}.}
\label{fig:vert}
\end{figure*}
The magnetoconvection at the solar surface distorts the magnetic field. Photospheric motions such as divergence, shear, and rotation tangle the magnetic field lines. This leads to a Poynting flux into the upper Solar atmosphere.
Figure \ref{fig:poynt} shows a volume rendering of the axial component of the Poynting flux entering the loop from the photospheric layer at each footpoint.
The resulting Poynting flux has a complex structure of oppositely directed strands in close proximity. The Poynting flux varies on short timescales. In our model we find that the axial component of the Poynting flux associated with transverse flows  dominates over the contribution by vertical motions carrying the horizontal magnetic field. The entanglement of strands of Poynting flux suggests braiding of the magnetic field.

The distribution of temperature and density across the loop is highly inhomogeneous. 
Fig. \ref{fig:swirl_cuts} and Fig. \ref{fig:vert} show cuts through the simulation box. 
Fig. \ref{fig:swirl_cuts} reveals a complex structure of the loop cross-section at the loop apex. 

Several regions of enhanced Poynting flux are visible in panel (f) of Fig. \ref{fig:swirl_cuts}. A complex structure of strong thin current sheets has formed in the loop \footnote{The unit abampere per square centimeter ($\mathrm{biot\; cm^{-2}}$) corresponds to $1\times 10^{5} \mathrm{A\; m^{-2}}$ in SI units and to $c\;\mathrm{stA\; cm^{-2}}$ in the CGS-ESU system, with $c$ being the speed of light. (see panel (g)) in Fig. \ref{fig:swirl_cuts}). The regions with increased Poynting flux show enhanced temperature. The location of the hottest plasma filaments does not exactly coincide with  the highest plasma densities, but the hot regions show enhanced densities.} This is consistent with the well-known scaling relations \cite[for example][]{1978ApJ...220..643R} that predict higher temperature and density in the case of increased energy flux into the loop \cite[for example Eqs. 3 and 4 of][]{2012A&A...537A.152P}.
Tracing back magnetic field lines from the loop apex shows that the hottest regions are connected to the interior of magnetic concentrations.

In our model, we find the loop to form numerous strands all along the loop, creating clearly discernible substructures. This is in contrast to \citet{2016ApJ...830...21R} who find only one smooth homogeneous hot and dense structure in their numerical model.
Instead, in our model the loop consists of individual filaments exhibiting a wide range of temperatures in the midplane from 0.9-4 MK and electron densities from $\mathrm{1\times 10^{8}\; cm^{-3}}$ to $\mathrm{5\times 10^{8}\; cm^{-3}}$. The top panel of Fig. \ref{fig:vert} shows that the temperature is enhanced in numerous strands spanning the entire loop length. Likewise, the electron density is increased in structures with a width of several hundred kilometers in the coronal part of the loop. We speculate that we find substructures in our simulation because of the significantly lower diffusivity of our numerical model compared to \citet{2016ApJ...830...21R}. Alternatively, it might be that \citet{2016ApJ...830...21R} do not find substructures in the loop because they apply a smooth driving at the footpoints.

Up- and downflows occur within the loop structure at the same time along different strands within the loop. Figure \ref{fig:swirl_cuts} e) shows two regions with oppositely-directed flows in the midplane, seen clearly as patches colored in red and blue. In each region the plasma is moving predominantly in one direction along the loop axis.
As illustrated in Fig. \ref{fig:vert}, we find cool and dense upflows with a temperature around $10^{4}$ K into the corona with a width of several hundred kilometers and velocities of 10 to 30 $\mathrm{km\;s^{-1}}$ while low-density plasma in the coronal part of the loop at temperatures at and above 1 MK can reach axial velocities of over 50 $\mathrm{km\;s^{-1}}$. Structures of cool plasma extend up to several megameters above the solar surface. These jets, however, are too slow to be type \RomanNumeralCaps 2 spicules \citep{2007PASJ...59S.655D}. The upflow velocity, however, is compatible with typical speeds of type \RomanNumeralCaps 1 spicules of 10-20 $\mathrm{km\;s
^{-1}}$. We do not expect to accurately simulate Type \RomanNumeralCaps 2 spicules due to the low resolution of the chromosphere and transition region. Additional physics may also be required; for example, ambipolar diffusion is necessary to generate spicules in simulations with the Bifrost code \citep{2017Sci...356.1269M}.

\subsubsection{Current sheets and heating rate}
\label{section:currents}

\begin{figure*} 
\resizebox{\hsize}{!}{\includegraphics{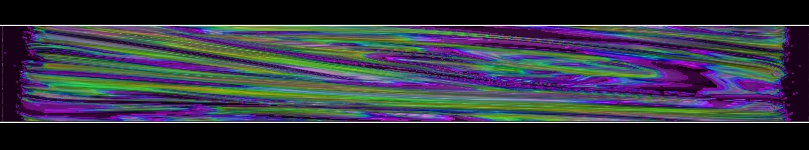}}
\caption{Volume rendering of the current density squared in the computational box. The range of the color scale is from $4.8\times 10^{-24}$ to $8\times 10^{-17}\; {\rm{biot}}^{2}\;{\rm{cm}}^{-4}$. See Sect. \ref{section:currents}.}
\label{fig:volrenj}
\end{figure*}

\begin{figure*} 
\sidecaption
\includegraphics[width=.6\textwidth]{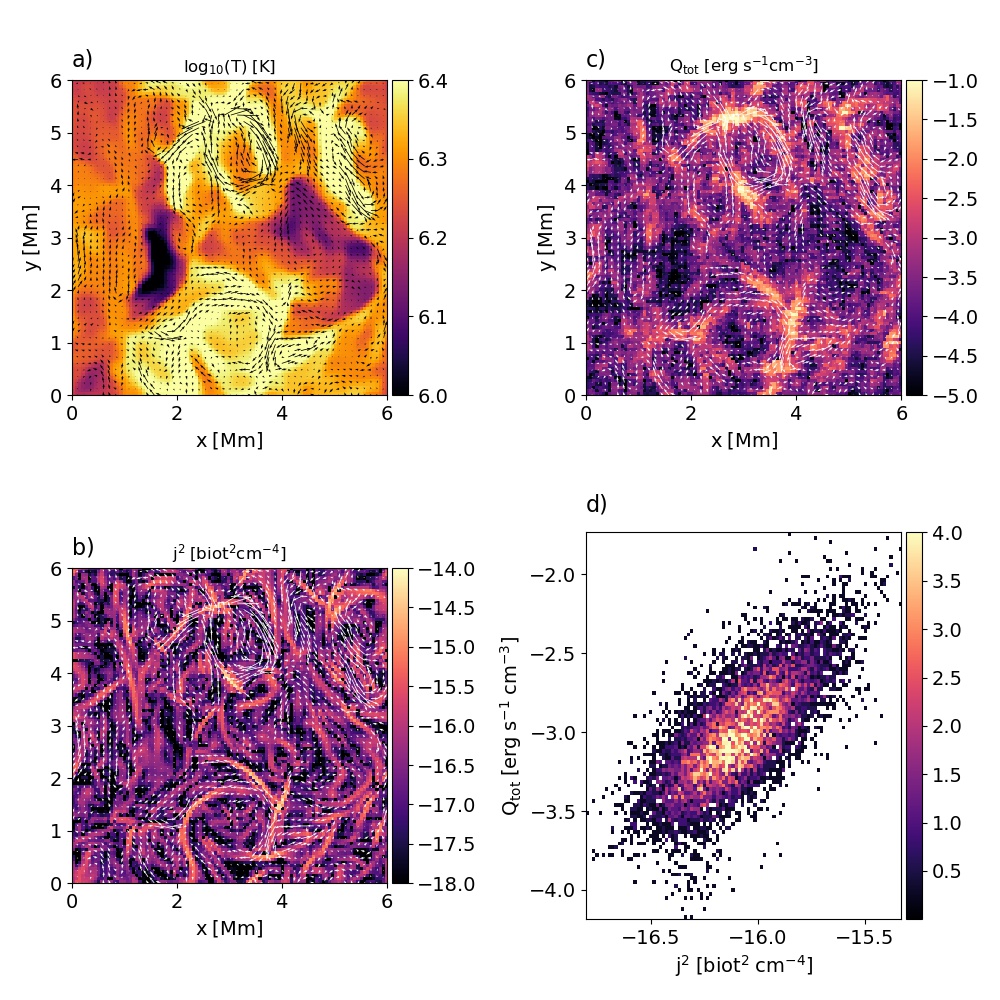}
\caption{Cut through the loop at the apex (cross section perpendicular to the loop axis). (a) Temperature; (b) Sum of viscous and resistive heating; (c) Squared current density; (d) 2D histogram of the total numerical heating rate vs. the squared current density.}
\label{fig:heating}
\end{figure*}

The current sheets resulting from the footpoint motions are elongated along the loop axis (see Fig. \ref{fig:volrenj}). Many of the current sheets are stretched all along the loop, roughly aligned with the magnetic field.  This is similar to the current sheets in the straightened-loop simulations by \citet{2017ApJ...844...87R} and the simulation of a stable active region by \citep{2011A&A...530A.112B}. Due to the approximate alignment with the magnetic field, the inclination of the current sheets points to fieldline braiding. In agreement with findings of those earlier studies, we see that the temperature profile follows predominantly the concentrations of the current.

The relation between the heating rate arising from numerical diffusivity and current sheets is nontrivial. A comparison between the numerical heating and the current sheets computed from the magnetic field is shown in Fig. \ref{fig:heating}. The cuts through the x-y-plane in panels (a) and (b) show that clearly the temperature is enhanced where most energy is deposited. The energy dissipation term is numerical and tailored to act only in locations with the largest gradients, as we pointed out in Sect. 2.1 and as described in more detail by \citet{2017ApJ...834...10R}. Here we show the total (numerical) heating rate, that is, viscous plus resistive. The spatial distribution of these two components is very similar, with (for this snapshot) the viscous heating being about a factor of 5.4 stronger than the resistive term in the coronal part. With the heating being defined through a numerical procedure, we cannot expect an exact one-to-one correlation to the currents, or more precisely the currents squared. Still, comparing panels (b) and (c) we see that mostly threads of enhanced currents and enhanced (numerical) heating rate coincide. This is underlined by panel (d)  that shows that there is a correlation between the two quantities, albeit with significant scatter.\\
The evolution of the loop is dynamic, current sheets continuously form and dissipate. Variations can be seen on a timescale down to seconds, though not shown in Fig. \ref{fig:avg}. 
While the current sheets evolve on a short timescale, the evolution of the heated plasma filaments occurs at a slower rate. The evolution of the temperature of individual heated filaments occurs on time scales $\geq$5 min because of the longer cooling time scales through heat conduction and radiation.

The location of the current sheets is associated with strong horizontal flows as can be seen in panels (c) and (f) of Fig. \ref{fig:swirl_cuts}.
We find the formation of multiple small-scale current sheets, which are continuously formed and fragment into smaller parts. The observed range of scales of the current sheets extends down to the dissipation scale, indicating turbulent behavior \citep{2016ApJ...817....5H}. 

The substructure of the loop is often organized in swirls. These are best seen through the velocity component perpendicular to the loop. In Fig. \ref{fig:swirl_cuts} c we show a cut at the loop apex that shows one prominent swirl (highlighted by the box).
Swirling motions in the lower atmosphere due to rotational motions at the solar photosphere have been reported in observations and simulations. They have been related to heating in locations of shear flows \cite[e.g., ][]{2012Natur.486..505W,2014PASJ...66S..10W, 2011A&A...533A.126M,2012A&A...541A..68M,2021A&A...645A...3Y}. Here we now see such swirls to extend all the way into the corona where they can still be found at the apex of a coronal loop at a height above the photosphere of 16 Mm. A detailed analysis of these structures will be presented in a subsequent paper.

Likewise, we find these swirls to show an increased temperature as illustrated in Fig. \ref{fig:swirl_cuts} a which shows a rotating structure with a width of 2 Mm in the loop midplane (see yellow box). Current density, temperature, and Poynting flux are all enhanced at the outer edge of the swirl (panels (a), (e), (g)). Currents building up at the edges of flux tubes due to the shear caused by the velocity gradient were studied before in \citet{2006A&A...451.1101D}, but in our model these swirls are driven self-consistently by the near-surface magnetoconvection. The rotating swirl is also clearly visible in the EUV emission at 211 \AA\ and in X-rays (see Fig.\,\ref{fig:swirl_cuts}d and h. While it appears bright in X-rays, in the EUV it is darker than the surrounding loop. This is also the case for the other swirls in the loop cross-section. This is because in this swirl the temperature exceeds 2 MK (Fig 8a) and it thus is too hot to emit in the 211 \AA\ channel of AIA. With X-rays being sensitive to higher temperatures, these then show a significant brightening. The role these vortices play for coronal loop heating will be subject to a separate study.

\begin{figure*} 

\resizebox{\hsize}{!}{\includegraphics{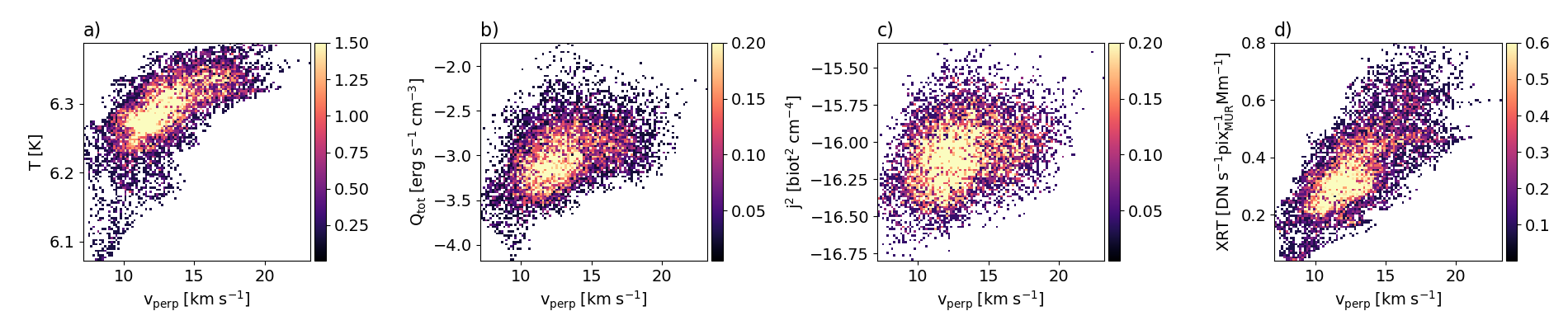}}
\caption{2D histograms of the temperature, total heating, squared current density, and X-ray emission at the apex vs. the velocity perpendicular to the loop axis. The X-ray emission has been computed to correspond to what XRT on Hinode would measure.
The quantities have been averaged over a time range of 15 minutes. See Sect. \ref{section:currents}.}
\label{fig:histo}
\end{figure*}

Motions perpendicular to the magnetic field are mainly responsible for carrying energy upward into the corona. In our model we find that this $s$-component of the Poynting flux (along the loop axis) dominates the contribution by vertical motions carrying the horizontal magnetic field (in the $x$ and $y$ directions). Rotating motions that extend upward in nearly vertical vortices are common in the intergranular lanes, but other driving motions such as shear motions are equally possible. Figure \ref{fig:histo} shows a 2D histogram of the temperature, total heating rate, and squared current density at the loop apex as a function of the velocity perpendicular to the loop axis. This is generally in agreement with \cite{2020ApJ...894L..17Y, 2021A&A...645A...3Y}, although their simulations do not reach into the corona. The total heating rate is the sum of resistive and viscous heating.
An increase of the heat input results in an immediate temperature increase and thus an increase of coronal emission. 
Consequently, we find a positive correlation between the coronal emission and the velocities perpendicular to the loop.
We find that increased heating rates, current densities, and plasma temperatures are clearly associated with 
higher perpendicular velocities. The temperature is enhanced at the location of current sheets created by velocity shear. What one has to keep in mind is that due to the low plasma beta of $7\times 10^{-3}$ the Lorentz force will dominate over the gas pressure and the plasma dynamics in the corona are driven by the magnetic field, which follows vortical photospheric flows, not the other way around.
The correlation we find between emission and  LOS-velocity (panel (d)) in Fig.\,\ref{fig:histo}) is consistent with the observed correlation between line intensity and nonthermal broadening \cite[e.g.,][]{2000A&A...360..761P,2015ApJ...799L..12D}. 
We do not see an instantaneous correlation between a peak in the heating rate and the density (Figs.~\ref{fig:spacetime} and Fig. \ref{fig:apex}). A temperature increase, however, leads to chromospheric evaporation and thus a delayed increase of the loop density. The data have been averaged over 15 minutes of solar time before computing the histograms, so that delayed effects of the heating are also taken into account. The parts of the loop with increased heating are eventually filled with higher-density plasma and brighten in the X-ray band.

\subsubsection{Isotropy of the velocity amplitude}
The velocity components perpendicular to the loop axis have different properties compared to the velocity parallel to the loop axis. The perpendicular velocity is structured on small scales like the currents and is driven by the turbulent motions of the magnetic field. The parallel velocity is structured on larger scales and driven in response to the evaporation of plasma along the guide field due to heating. The distribution of the magnitude of the velocity components, however, is nearly isotropic. We find velocities in the corona perpendicular to the loop axis with an amplitude of up to 80 $\mathrm{km\;s^{-1}}$.
The flows along the loop axis have a similar amplitude. However, comparing the spatial structuring of the perpendicular and the axial flows (Fig.~\ref{fig:swirl_cuts}c and e reveals that in general the axial flows are organized on larger spatial scales (across the loop) ranging from several hundred kilometers up to 2 Mm.

This is in contrast to \citet{2020A&A...639A..21P}, whose loop models lack the dynamics along the guide field because they do not include the interaction with the chromosphere and thus evaporation of plasma that drives the parallel flows. Hence, \citet{2020A&A...639A..21P} find a strong deviation from isotropy with the axial speeds being much smaller than the flow speeds perpendicular to the loop axis.

\subsection{Synthesized emission}
\label{subsection:synth}

\begin{figure*} 
\resizebox{\hsize}{!}{\includegraphics{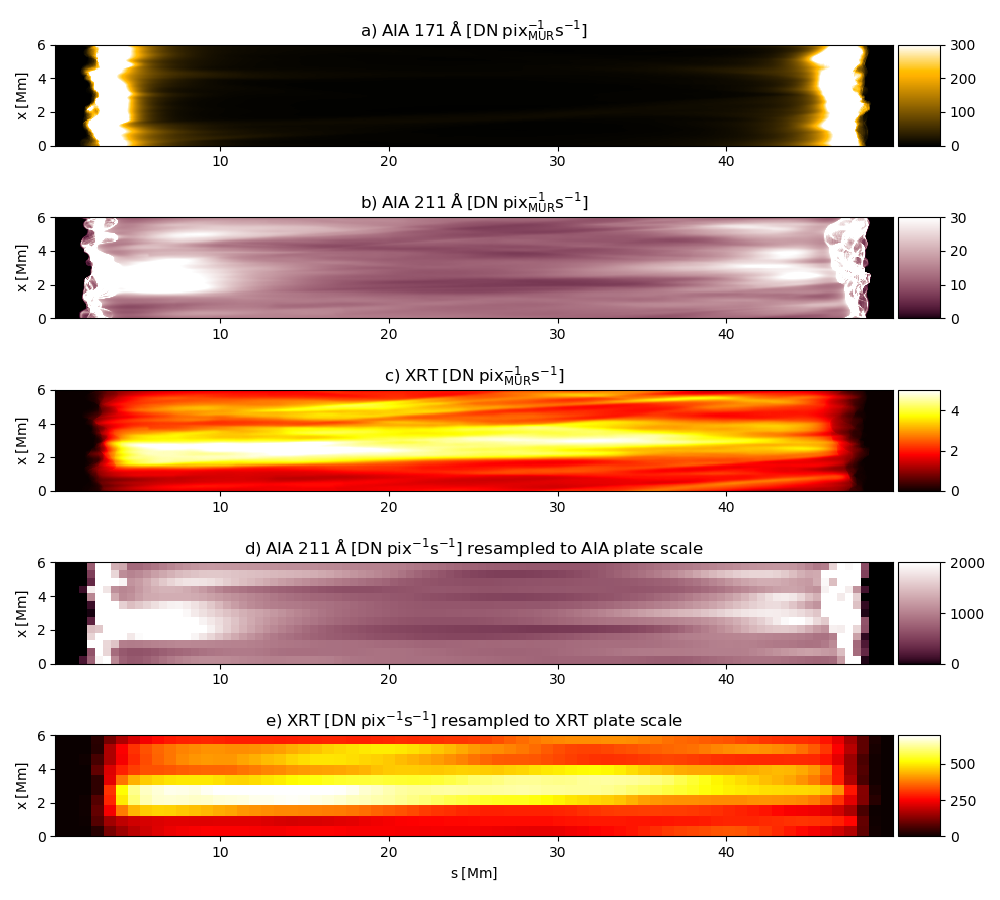}}
\caption{View of the simulated loop from the side as seen in EUV and X-ray observations. Panels (a) to (c) show the view at the original resolution of the numerical model, panels (d) and (e) the images downsampled to the plate scale of the instruments. From top to bottom: Synthetic emission for the 171 \AA\; and 211 \AA\; AIA bands integrated along the y-axis, emission as would be measured with the Al-poly filter of XRT, degraded synthetic emission in the 211 \AA\; band taking into account the pixel size of the AIA instrument of $0.6^{\prime\prime}$, degraded emission for XRT with a pixel size of $\sim 1^{\prime\prime}$. The synthetic emission has been integrated along the line of sight perpendicular to the loop axis.
See Sect.\,\ref{subsection:synth}.}
\label{fig:em}
\end{figure*}

\begin{figure*}
\sidecaption
\includegraphics[width=.7\textwidth]{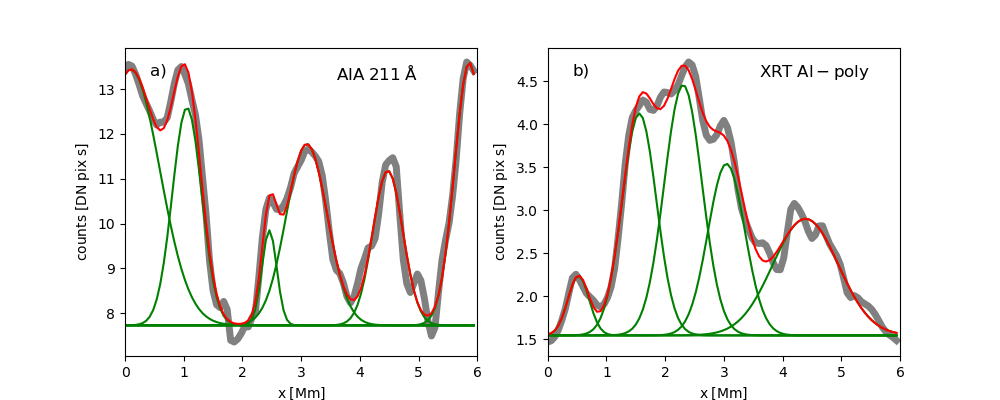}
\caption{Cross-sectional cut of the loop in coronal emission.
The gray thick lines show cuts at the apex of the loop as displayed in Fig.\,\ref{fig:em} for AIA 211\,\AA\ and XRT at the original resolution.
To quantify these cross-sectional cuts we show multi-Gaussian fits with a constant background in each case.
The individual Gaussians of the fit are shown in green, the sum of all Gaussians in red.
See Sect.\,\ref{subsection:synth}.
}
\label{fig:strands}
\end{figure*}

To compare our loop model directly to actual observations by, for example, AIA or XRT we derived the emission from the model as these instruments would observe it. For this we employed the procedure described in Sect. \ref{section:em}.
 The average loop temperature ranges from 1.3 MK to 1.8 MK as can be seen in Fig. \ref{fig:avg}. Consequently, the emission corresponding to the AIA 171 \AA\ channel is very much concentrated at the footpoints, with negligible emission in the upper part of the loop compared to the footpoints as shown in panel (a) of Fig. \ref{fig:em}. This is very similar to the moss emission observed at the footpoints of hot loops \cite[e.g.,][]{1999SoPh..190..419D, 1999SoPh..190..409B}. In this sense the loop we study here represents a hot loop with moss emission at its footpoints. Consequently, to investigate the substructure of the loop in the corona we employed an AIA channel showing hotter plasma and X-rays. For the AIA instrument we chose to focus on the 211 \AA\; band, which is sensitive to emission from plasma with a temperature near 2 MK. 
 
The plasma in our simulation reaches temperatures of three million Kelvin or more in the coronal part of the simulation domain, with the peak temperature being 5.2 MK.
Plasma at these temperatures emits mainly in the X-ray part of the spectrum. The synthesized emission for the 211 \AA\; band and the X-ray emission is shown in Fig. \ref{fig:em}b and c, respectively. The emission in both wavelength bands displays a clear substructure with several bright slender strands running almost from footpoint to footpoint.

The emission computed in the AIA bands is significantly stronger near the footpoints where the plasma is colder and denser than in the upper atmosphere. The AIA 211\,\AA\ emission near the footpoints shows multiple bright thin features that show heated dense low-lying plasma. A future study will have to show to what extent these are low-lying features separate from the coronal part of the loop, or if these are actually a signature of small-scale heating events in the low part of the loop that contribute (significantly) to the energization of the corona as a whole.

Discrete strands are visible both in the 211 \AA\; channel and in X-ray. To illustrate the fine structure of the loop emission, Fig. \ref{fig:strands} shows a cut through the line-of-sight integrated emission for both filters. 
We applied a multi-Gaussian fit to the cross-sectional cuts to determine the full width at half maximum (FWHM) of the strands. This FWHM in the range of of 290 to 1200 km for the strands in the 211 \AA\ channel and 370 to 1300 km for the strands in X-ray emission. This is consistent with typical strand widths observed by Hi-C \cite[e.g.,][]{2013ApJ...772L..19B,2020ApJ...892..134W}. Also, the finding that the structures appear broader at higher temperature is consistent with observations that show structures at higher temperatures to become more fuzzy \cite[e.g.,][]{2009ApJ...694.1256T}. The lifetime of the strands is 3 to 20 minutes. 

When comparing the artificial emission to observations, we need to take into account the pixel sizes of the instruments, which differ from the pixel size of our simulation. The AIA instrument has a plate scale of $0.6^{\prime\prime}$ per pixel (and a resolution of about $1.4^{\prime\prime}$). On the Sun near disk center this corresponds to a pixel size of 450 km. For XRT the plate scale is $1^{\prime\prime}$ per pixel (corresponding to 725 km) at a resolution of about $2^{\prime\prime}$. We resampled the synthesized emission to the instrument plate scale. For simplicity we did not convolve with  the point spread function, which should be sufficient here for illustrative purposes. The count rates for a patch of neighboring pixels of the numerical model, which correspond to the instrument plate scale, were summed up. While individual strands are still visible in parts of the loop for AIA, discerning different strands is more difficult for XRT which has a coarser resolution. The resampled synthesized emission is displayed in panels (d) and (e) of Fig. \ref{fig:em}. For XRT the emission shows two main structures, a bright wide strand and a dimmer narrow strand. For the decreased resolution, only the widest strand is clearly distinguished from the background, while the dimmer strand is only visible in parts of the loop and appears to be partially merged with the larger strand. For the emission in the 211 \AA\ channel, the four main strands that appear in the undegraded emission are still distinguishable, although not along the whole length of the loop. The substructure on scales of a few hundred kilometers that is shown in the perpendicular cut in Fig. \ref{fig:swirl_cuts}, is not resolved.

\section{Discussion}
\label{section:disc}

In this section we cover the source of the energy injection into the corona in our simulation, the formation of current sheets in the loop, the relation between the photospheric driver and flows in the hot corona, energy input by global loop oscillations, and, finally, predictions for observation of the simulated loop with AIA and XRT. We compare our results to previous models of straightened loops and observations.

\subsection{Poynting flux and heating}
\label{subsection:poynt_heat}

Convective motions lead to the deformation of magnetic concentrations that twist and shear the magnetic field.
As the main source of the heating we find photospheric motions within elements of strong magnetic flux concentrated in the intergranular downflow lanes, rather than braiding due to horizontal motions of magnetic features relative to each other. We observe rotational motions of the plasma within the magnetic concentrations that lead to rotation and twisting of traced magnetic field lines. The small-scale horizontal motions within the magnetic patches and the deformation of the patches lead to an upward directed Poynting flux into the loop. The magnetic field lines connected to hot patches in the corona are mainly anchored in parts of the intergranular downflow lanes which show both strong magnetic fields and increased horizontal flows (see, for instance, the footpoint of field line in Fig. \ref{fig:photo}). Rotational motions have also been found in simulations by \citet{2012Natur.486..505W,2012A&A...541A..68M,2013ApJ...776L...4S,2020arXiv200413996Y,2021A&A...649A.121B}.
The Poynting flux  is injected into the coronal loop from both footpoints. It is dissipated throughout the corona without a preferred location in the coronal loop.

We find a "bursty" heating profile along individual field lines that strongly varies on short timescales and small spatial scales (see Sect. \ref{section:ftrace}). Heating events are of short duration and distributed throughout the atmosphere. While heating can occur along the full length of a field line, strong heating events appear to be more localized. Both Poynting flux and heating rate fluctuate strongly in space and time. We find that strong Poynting flux events at a point along the field line (here chosen to be the apex) are associated with or closely followed by an increase in the heating rate and a delayed increase in density along the studied fieldline, consistent with chromospheric evaporation (see Fig. \ref{fig:apex}). 

Peaks in the total heating rate are superimposed on continuous background heating. \citet{2011A&A...530A.112B} investigate energy input and heating along individual field lines in their 3D active region simulation. They find heating predominantly near the footpoints for fieldlines connecting the main polarities of the spots in the simulated active region to network due to the higher velocity shear at the footpoints located in weaker magnetic field concentrations, but also find field lines that show strong heating events at the loop top consisting of several short small-scale events. In our simulation, we find no clear concentration of the heating near the footpoints.
\citet{2013A&A...550A..30B} investigate the temporal statistics of heating events and find short spikes below a minute imposed on slower variations in individual gridpoints, which is consistent with our results. \citet{2017A&A...603A..83K}
also studied the size distribution of the heating events and found that large energy release events are favored, which is in contrast to our findings, which show many small events. Their algorithm used to  identify heating events, however, cannot resolve clusters of small heating events that are closely packed, thus the number of small heating events might be underestimated in their analysis. 

\subsection{Current sheets}

The structure of the current density in our model qualitatively resembles the current sheets found by \citet{2018A&A...615A..84R} as a result of a magnetic avalanche. \citet{2018A&A...615A..84R} study the disruption of three magnetic threads after undergoing an initial instability. The end state of the simulation after the onset of the Kink instability and the magnetic avalanche is a braided system that undergoes continuous dissipation. The current sheets then form a network of small complex structures enabling reconnection of magnetic field lines and subsequent heating \citep{2020A&A...633A.158R}.

The resistivity in numerical simulations is several orders of magnitude higher than on the real sun. This leads to an onset of reconnection at smaller misalignment angles between magnetic field lines than would occur in the plasma on the sun. This problem is avoided in relaxation studies by starting from a prescribed braided flux tube with large misalignment angles of the field lines as initial condition. In contrast to relaxation models such as the simulations conducted by \citet{2010A&A...516A...5W} and \citet{2011A&A...525A..57P}, we do not start with an artificially braided field as an initial condition that undergoes an instability and relaxes to a new equilibrium. Instead, the system is in a continuously driven state due to photospheric motions. 

\citet{Rappazzo_2008} conducted simulations of a coronal loop in Cartesian coordinates in which slow photospheric motions drive a Poynting flux, leading to a turbulent cascade of energy from the scale of convective motions to small scales where the energy is dissipated. They find that the field lines are barely entangled, current sheets still form and are continuously dissipated. This is similar to the situation we find in our simulations. The original Parker model predicts angles of 20\textdegree\; between braided field lines, which is not observed \citep{van_Ballegooijen_2011}.
We find an inclination angle of the magnetic field in the range of 0-70\textdegree\; in the coronal part of the loop with a mean value of 5.4\textdegree\;. This is consistent with previous simulations. 
Thus, reconnecting field lines in the coronal part of the computational domain are almost parallel. These angles are much smaller than the value of 30\textdegree\; predicted by nanoflare theories and not compatible with the large misalignment angles required for the secondary instability to set in \citep{2005ApJ...622.1191D}.

\subsection{Atmospheric coupling}

An unanswered question is how the magnetic structure at the loop footpoints affects the internal structure of the loop.
One possibility is that the topology of the photospheric driver strongly influences the dynamics of the loop \citep{Rappazzo_2010}.
Electric currents should develop along field lines with footpoints that are subject to shear motion. The magnetic field would then be a mapping of the velocity pattern driving the field line tangling.
As found in \citet{2007ApJ...657L..47R} and \citet{Rappazzo_2008, Rappazzo_2010} the magnetic field lines have a topology mostly independent of the photospheric velocities.
\citet{Rappazzo_2008, Rappazzo_2010} argue that the formation and dissipation of current sheets arises from the nonlinear nature of the system and does not require complex footpoint motions of the magnetic field lines.
\citet{Ritchie_2016}, however, argue that the nature of the energy release does depend on the complexity of the photospheric driver, with coherent motions leading to a smaller number of large events and more complex drivers causing frequent small heating events. We find both imprints of the photospheric motions such as vortices in our simulations as well as turbulent flows not directly related to footpoint motions. Therefore, both statements are not mutually exclusive.

Although temperature and density are generally enhanced in the region of the simulation box above the strongest magnetic field concentrations in the photosphere, in our model the distribution of temperatures in the corona does not follow directly the distribution of the magnetic field at the photosphere. Instead, the photospheric driving leads to a braided state showing aspects of turbulent behavior. 
The key role that MHD turbulence plays here is that it transports energy to smaller scales where it is dissipated \citep{Rappazzo_2008}, for example by creating small-scale local field reversals \citep{2020arXiv200406186J}. 

Care has to be taken when following individual field lines in a turbulent flow.
The algorithm can jump between different field lines due to numerical errors and becomes structurally unstable in the vicinity of reconnection regions. It is therefore not possible without a doubt to distinguish between field line jumps due to physical reconnection events and jumps due to numerical inaccuracies.
As illustrated in Fig. \ref{fig:spacetime}, changes of identity of a field line can be identified by a sudden change of plasma parameters. In between those events, however, tracing of a field line is reasonably reliable, although the output of the field line tracing algorithm should be checked manually.

\subsection{Emission}

In agreement with \citet{2017ApJ...837..108P}, we find that the synthesized emission does not necessarily show a braided appearance despite the braided state of the magnetic field.
The bright strands in the 211 \AA\ channel have a smooth appearance and do not show clear indications of overlapping strands. This is not necessarily a sign, however, that no braiding of loop strands occurs \citep{2017ApJ...837..108P}. Especially in the hot part of the loop, evidence for the crossing of strands is visible in some snapshots in the X-ray emission. 

The discrete strands visible in the emission are in contrast to the simulation by \citet{2016ApJ...830...21R} which yields a loop that appears monolithic. The appearance of a substructure might be due to the significantly lower diffusivity of our numerical model. We find an effective resistive diffusivity of $\mathrm{2.4\times 10^{12}}\; \mathrm{cm^{2}s^{-1}}$ in the corona (see Sect. 3.3).

The width and lifetime of the strands that we see in the synthesized observations is compatible with observed threads in coronal loops \citep{2013ApJ...772L..19B,2020ApJ...892..134W,2020ApJ...902...90W}. Strand widths range from 250 to 1400 km FWHM for the AIA 211 \AA\ channel and 600 to 1500 km FWHM for the XRT channel. \citet{2013ApJ...772L..19B} examine a set of 91 loops observed by Hi-C and find strand widths from 212 km FWHM to 1665 km FWHM, thus the strand widths we find in our simulation are within the observed range for both channels. \citet{2020ApJ...902...90W} find that most of the strands have widths between 200-760 km FWHM observed by Hi-C 2.1. The lifetimes that we find for individual coronal loop strands is on the order of minutes, which is consistent with observed temporal variablity of loop emission, although observed loops (in which, most probably, the strands are not resolved) usually remain bright as a whole for several hours \citep{2014LRSP...11....4R}. 

If the synthesized emission is degraded to the instrument resolution of AIA and XRT, the loop has a mostly smooth appearance. To detect the substructure, a higher resolution would be needed. The Hi-C instrument \citep{2014SoPh..289.4393K} and the EUI instrument \citep{2020A&A...642A...8R} on Solar Orbiter have a high enough spatial resolution to detect strands of the width found in our simulations.
 The emission in the 211 \AA\ channel exhibits more fine structuring than the emission from hotter plasma in X-ray which has a smoother appearance, consistent with observations \citep{2009ApJ...694.1256T}. 

A cut perpendicular to the axial direction, as shown in Fig. \ref{fig:swirl_cuts}, reveals a highly complex structure of the density and temperature distribution throughout the loop cross-section, the strands seen in the emission after integration along the line of sight might therefore in part be caused by projection effects. 
\citet{2017ApJ...837..108P} also find a nonzero angle between magnetic field lines and bright strands in regions exhibiting frequent reconnection events which cause changes of the connectivity on a timescale on the order of the timescale of heat conduction parallel to the field lines.

\section{Conclusions}
\label{section:conclusion}

In this paper we investigated how the energy to heat a coronal loop is generated in the photosphere, injected along the magnetic field, and finally dissipated.
We tailored the 3D computational domain to contain a single coronal loop.
Essentially, this is a straightened loop, with gravity pointing downward
on both ends and decreasing to zero at the apex of the loop.
This setup allows us to study the details within the loop on scales not achievable in a 3D model covering a whole active region. What distinguishes our model from previous setups that have been similar in geometry \cite[e.g.,][]{2016ApJ...830...21R, Rappazzo_2008} is the realistic treatment of the bottom boundary at the solar surface (at each of the ends of the loop).
Here we fully account for the near-surface magneto-convection that self-consistently drives the evolution of the photosphere where the loop is rooted.
With this model setup, we can investigate the formation and evolution of (sub)structures within the loop and how they are driven by the motions of magnetic patches
at the surface. 
As a response to the driving, a complex fine structure of small-scale current sheets and flows develops in the loop.
In its coronal part, the loop plasma exhibits turbulent behavior, for instance, visualized through the spatial structures of the current and the velocity in cuts across the loop.
The temperature distribution higher up in the loop, however, does not directly follow the magnetic field distribution at the photosphere in a simple fashion.
Energy is supplied to the upper atmosphere via the Poynting flux originating from the photospheric layer, and we can follow the Poynting flux as it propagates along individual field lines.
At the same time, we see frequent reconnection of fieldlines, illustrated through sudden changes of, for example, density.
The energy input, intermittent in both space and time, results in a substructure of the loop seen in synthesized EUV and X-ray emission.

In our study we showed the power of loop-in-a-box models with a realistic photosphere driven by magneto-convection.
After having proven the concept, one of our next steps will be to conduct a detailed analysis of the (magnetic) coupling from the photosphere to  the hot corona, and of the photospheric roots of the individual strands seen in emission.
Another topic will be an in-depth analysis of the (an)isotropy of the velocity field that would be uncovered in real observations through an analysis of emission line widths. 
We find indications for oscillations and waves in the simulations, which we also plan to study more closely. 
With these types of studies we will be able to improve the  quantification of contributions of different heating mechanisms, such as field-line braiding, wave heating, or low-lying reconnection.
Future tests of our model simulations may be expected from the Solar Orbiter mission \citep{2020A&A...642A...1M}, in particular from the High Resolution Imager of the Extreme Ultraviolet Imager (EUI, \citet{2020A&A...642A...8R}) combined with the magnetic field measurments made by the High Resolution Telescope of the Polarimetric and Helioseismic Imager (PHI, \citet{2020A&A...642A..11S}).

Improved future simulations would profit from a significantly smaller grid size. Also, the current simulations suffer from the treatment of radiation in the chromosphere in LTE. This shortcoming will eventually be addressed by using the new version of MURaM \citep{Przybylski_2021}.

\begin{acknowledgements}

The authors would like to thank the referee, Oskar Steiner, for constructive comments helping to improve the paper. This research has made use of SunPy, an open source and free community-developed solar data analysis package written in Python \citep{sunpy_community2020}.
This project has received funding from the European Research Council (ERC) under the European Union’s Horizon 2020 research and innovation programme (grant agreement No. 695075). We gratefully acknowledge the computational resources provided by the Cobra supercomputer system of the Max Planck Computing and Data Facility (MPCDF) in Garching, Germany. This material is based upon work supported by the National Center for Atmospheric Research, which is a major facility sponsored by the National Science Foundation under Cooperative Agreement No. 1852977.

\end{acknowledgements}

\newpage

\bibliography{paper}
\bibliographystyle{aa}

\end{document}